\documentclass[journal]{IEEEtran}
\usepackage{amsfonts, bm}
\usepackage{amssymb}
\usepackage{color,graphicx}
\usepackage{cite}
\usepackage[cmex10]{amsmath}
\usepackage{amsopn}
\usepackage{multirow}
\usepackage{bigstrut}
\usepackage{microtype}
\usepackage{verbatim}
\usepackage{stfloats}
\usepackage[ruled]{algorithm2e}
\usepackage{epstopdf}
\usepackage{subfigure}

\IEEEoverridecommandlockouts

\def\gap{0.75ex}
\abovedisplayskip\gap
\belowdisplayskip\gap
\abovedisplayshortskip\gap
\belowdisplayshortskip\gap

\graphicspath{{figs/}}

\allowdisplaybreaks

\newtheorem{proposition}{Proposition}
\newtheorem{theorem}{Theorem}

\newtheorem{remark}{Remark}

\newtheorem{lemma}{Lemma}

\begin{document}
\title{{Information-Theoretic Limits of Integrated Sensing and Communication with Correlated Sensing and Channel States for Vehicular Networks}}
\author{Yao Liu, Min Li, An Liu, Jianmin Lu, and Tony Xiao Han~\thanks{Yao Liu, Min Li, and An Liu  are with the College of Information Science and Electronic Engineering, Zhejiang University, Hangzhou 310027, China (e-mail: \{yao.liu, min.li, anliu\}@zju.edu.cn). 	
Jianmin Lu and Tony Xiao Han are with Huawei Techologies Co., Ltd. (email: \{lujianmin,tony.hanxiao\}@huawei.com).
This work was supported in part by Huawei Techologies Co., Ltd. (\textit{Corresponding author: Min~Li.})
}}
\maketitle
\begin{abstract}
In connected vehicular networks, it is vital to have vehicular nodes that are capable of sensing about surrounding environments and exchanging messages with each other for automating and coordinating purpose. Towards this end, integrated sensing and communication (ISAC), combining both sensing and communication systems to jointly utilize their resources and to pursue mutual benefits, emerges as a new cost-effective solution.
In ISAC, the hardware and spectrum co-sharing leads to a fundamental tradeoff between sensing and communication performance, which is not well understood 
except for very simple cases with the same sensing and channel states, and perfect channel state information at the receiver (CSIR). In this paper, a general point-to-point ISAC model is proposed to account for the scenarios that the sensing state is different from but correlated with the channel state, and the CSIR is not necessarily perfect. For the model considered, the optimal tradeoff is characterized by a capacity-distortion function that quantifies the best communication rate for a given sensing distortion constraint requirement. 
An iterative algorithm is proposed to compute such tradeoff, and a few non-trivial examples are constructed to demonstrate the benefits of ISAC as compared to the separation-based approach.
\end{abstract}

\begin{IEEEkeywords}
	Integrated sensing and communication, connected vehicular networks, correlated sensing and channel states, capacity-distortion tradeoff.
\end{IEEEkeywords}

\section{Introduction}

\begin{figure}[!t]
	\centering
	\includegraphics[width=0.8\linewidth]{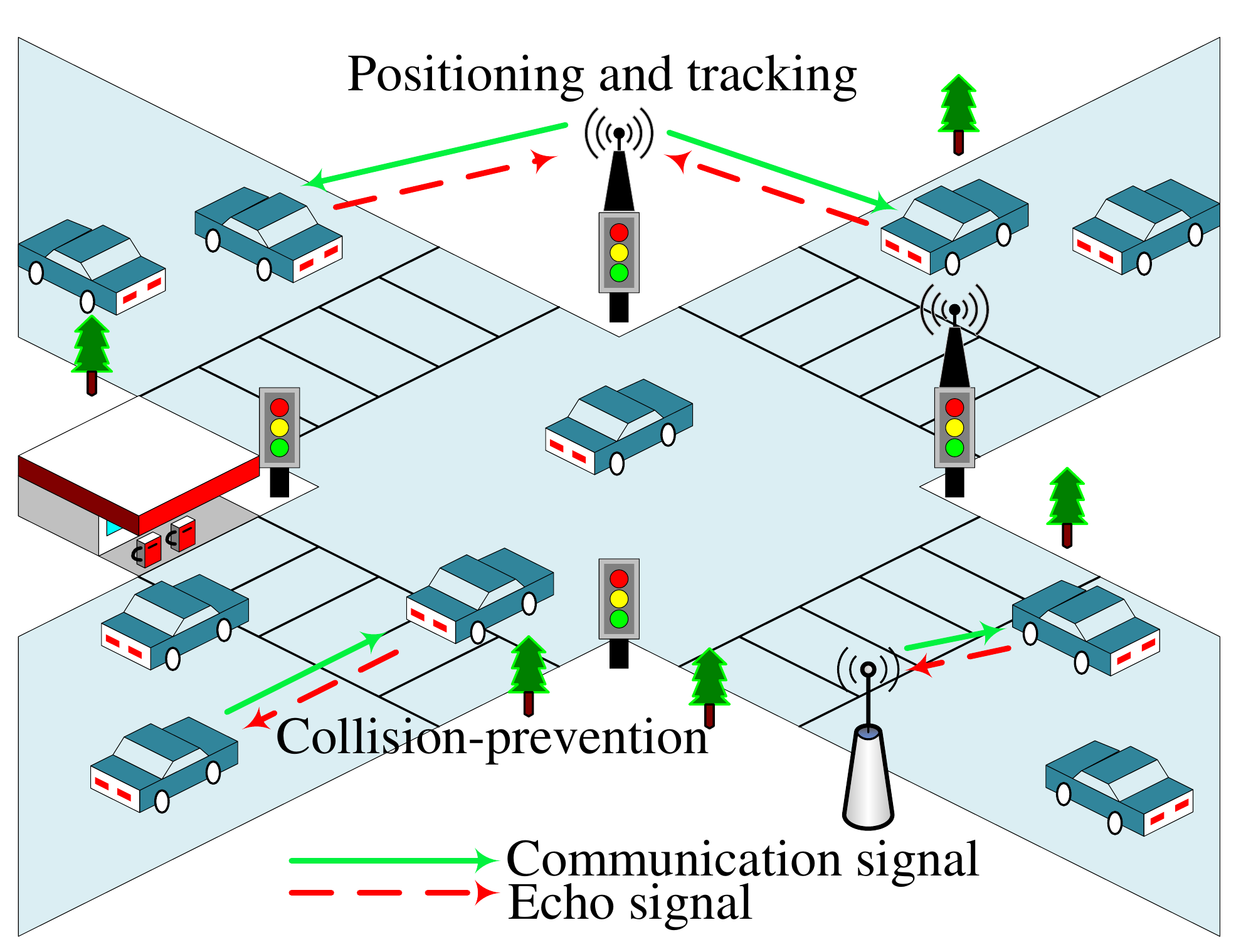}
	\caption{The applications of sensing and communication in vehicular networks.}
	\label{fig:vehicularNetworks}
	\vspace{0em}
\end{figure}

Connected vehicular network is expected to play an increasingly important role in future digital cities to support various applications such as autonomous driving and traffic management. As shown in Fig. \ref{fig:vehicularNetworks}, in addition to the dedicated communications that enable vehicular nodes to exchange messages with base stations and other vehicular nodes, high-accuracy sensing is also required for vehicular nodes to sense the environment and the states of surrounding vehicles (e.g., their speeds and locations) to perform collision-prevention and real-time traffic management. Integrated sensing and communication (ISAC) through co-sharing the same hardware 
and 
spectrum for both sensing and communication systems, is envisioned as an important new cost-efficient technology in such connected vehicular networks \cite{liu2022survey,wang2021joint}.

Substantial studies have been conducted to explore ISAC in different scenarios and system architectures, such as the state-of-art in the levels of coexistence, collaboration, and co-design \cite{liu2020joint}, the framework for large-scale mobile networks \cite{rahman2020enabling}, and the representative methodologies in vehicular networks \cite{ma2020joint}. 
There are also many studies focused on the key techniques of ISAC in vehicular networks, such as 
waveform design \cite{zhang2022integrated}, beamforming design \cite{mu2021integrated}, and spatial signal processing \cite{liu2020tutorial2}. While these studies demonstrate the advantages of joint design of sensing and communication, it is unclear whether these schemes have reached the optimal performance for the resources given. This motivates one to investigate the fundamental performance limits of ISAC and to quantify the optimal tradeoff between sensing accuracy and communication rate for a given ISAC scenario.

With the assumption of Gaussian parameters and estimation error, a notion of estimation information rate is introduced in \cite{bliss2014cooperative} to quantify the sensing accuracy, 
and the performance tradeoff bound is characterized in terms of estimation rate versus classic communication rate. With this approach, the performance limits for both single-antenna and multi-antenna scenarios are investigated in \cite{chiriyath2015inner, rong2018mimo}, respectively.
Instead of deriving equivalent estimation information rate for sensing, in \cite{kumari2017performance}, the authors propose to convert communication information rate to a mean-square error (MSE) metric and examine the tradeoff between the communication equivalent-MSE and the estimation MSE both in the same unit. 
However, this approach only works in a simple linear Gaussian modeling. 

To gain insights for more general ISAC systems, a notion of capacity-distortion function built on rate-distortion theory is introduced in \cite{kobayashi2018joint}. The sensing accuracy of parameters is quantified by general distortion functions while the communication performance is still quantified by classic communication rate. With the assumption of the same sensing and channel states and perfect channel state information (CSI) at the receiver (CSIR), the authors derive the optimal capacity-distortion function for point-to-point channels. More recent studies \cite{kobayashi2019joint, ahmadipour2021joint} move on to study the ISAC over multiple access channels (MACs) and broadcast channels (BCs), respectively. 

The aforementioned works provide important preliminary results on fundamental limit of ISAC systems. However, the strong assumption on sensing state and CSIR makes the model too restricted to capture the practical constraints for ISAC in vehicular networks. When performing collision-prevention, a car may wish to estimate the speed or position of the preceding car while communicating. Another circumstance is that the sensing targets may be part of the scatters in the surrounding environment of the car \cite{liu2020joint}. Therefore, the sensing state is usually different from but correlated with the communication channel state.
In addition, the CSIR can hardly be perfect due to the channel estimation
error. Given these practical constraints in vehicular networks, the modeling and the corresponding analysis of performance limits for more generalized ISAC systems are vital and imminent.


In this paper, we propose a more general point-to-point ISAC model, extending that of \cite{kobayashi2018joint}, to account for the scenarios that the sensing state at the transmitter is different from but correlated with the channel state, and the CSIR is not necessarily perfect. For the model studied, the fundamental limit of ISAC is characterized by a capacity-distortion function. The computation of the capacity-distortion tradeoff is further formulated as a constrained optimization problem, and an iterative algorithm based on Blahut-Arimoto algorithm \cite{blahut1972computation} is proposed to solve it efficiently.
A few non-trivial instances are provided to demonstrate the benefits of ISAC compared to the separation-based approach via time-sharing. 

\section{System Model} \label{Sec:systemModel}
\begin{figure}[!t]
	\centering
	\includegraphics[width=0.9\linewidth]{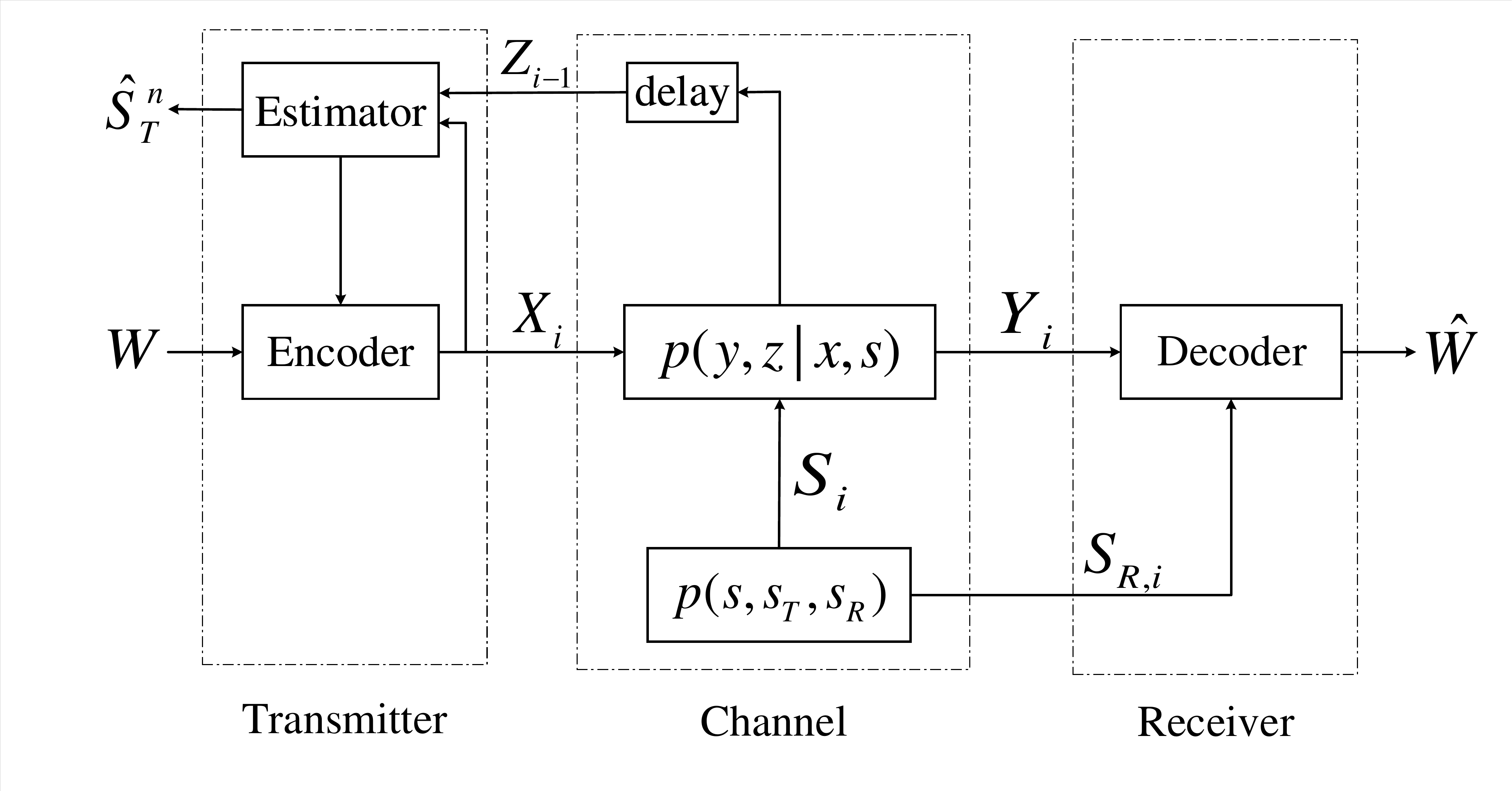}
	\caption{A general point-to-point ISAC channel with correlated sensing state and channel state, and imperfect CSIR.}
	\label{fig:channelmodel4p2pjcas}
	\vspace{0em}
\end{figure}

Consider a general point-to-point mono-static sensing ISAC model as shown in  Fig. \ref{fig:channelmodel4p2pjcas}. The transmitter wishes to convey a message $W\in[1:2^{nR}]\triangleq\{1,2,\cdots,2^{nR}\}$ to a decoder over a state-dependent discrete memoryless channel (DMC) while simultaneously estimating the sensing state sequence $S_T^n$ via
an output feedback. This feedback models the communication echo signals reflected back to the transmitter side, as in \cite{kobayashi2018joint,kobayashi2019joint, ahmadipour2021joint}. 
Mathematically, the state-dependent DMC considered is represented by tuple $(\mathcal{X},\mathcal{S},p(yz|xs),\mathcal{Y},\mathcal{Z})$ that consists of input alphabet $\mathcal{X}$, state alphabet $\mathcal{S}$, channel output alphabet $\mathcal{Y}$, feedback alphabet $\mathcal{Z}$, and a collection of conditional probability mass functions (PMFs) $p(yz|xs)$ for every pair of $(x,s)$. The channel state sequence $S^n$ is assumed i.i.d. according to $P_{S^n}\left({s^n}\right)=\prod\nolimits_{i=1}^n{P_S(s_i)}$. 
The CSIR is denoted by $S_{R}^n$, which is i.i.d. according to $P_{S_R^n}\left({s_R^n}\right)=\prod\nolimits_{i=1}^n{P_{S_R}(s_{R,i})}$ and the associated $S_{R}$ is assumed to be correlated with channel state $S$ but not necessarily the same. Such an assumption can encapsulate more general cases for CSIR: 1) $S_{R,i}=S_i$, the receiver has the perfect CSI; 2)  $S_{R,i}=\phi$, the receiver has no CSI; 3)  $S_{R,i}=g(S_i)$, the receiver has partial CSI. In this model, we also assume that sensing state sequence $S_T^n$ is i.i.d. according to $P_{S_T^n}\left({s_T^n}\right)=\prod\nolimits_{i=1}^n{P_{S_T}(s_{T,i})}$ and the associated $S_T$ is different from but correlated with the channel state $S$. 

For such a generalized model of ISAC, a $(2^{nR},n)$ code for the state-dependent channel consists of 
\begin{enumerate}
	\item a message set $\mathcal{W}=[1:2^{nR}]$;
	\item an encoder that assigns a symbol $x_i=f_i(w,z^{i-1})$ to each message $w\in\mathcal{W}$ and each delayed feedback $z^{i-1}\in\mathcal{Z}^{i-1}$ at time index $i$;
	\item a decoder that produces a message estimate $\hat{w}=h(y^n,s_R^n)\in\mathcal{W}$ upon receiving $y^n$ and observing $s_R^n$;
	\item a state estimator that assigns an estimated sensing sequence $\hat{s}_T^n\in\hat{\mathcal{S}}^n_T$ to each output feedback sequence $z^{n}\in\mathcal{Z}^{n}$ and the codeword $x^n\in\mathcal{X}^n$. 
\end{enumerate}

Similar to the studies \cite{kobayashi2018joint,kobayashi2019joint}, the sensing performance is measured by the expected distortion of the state estimated, i.e., 
\begin{equation}
	\mathbb{E}[d(S_T^n,\hat{S}_T^n)]=\frac{1}{n}\sum_{i=1}^n\mathbb{E}[d(S_{T,i},\hat{S}_{T,i})],\notag
\end{equation}
where $d:\mathcal{S}_T\times\hat{\mathcal{S}}_T\rightarrow[0,\infty)$ is a distortion function. A rate-distortion pair $(R,D)$ is said to be achievable if there exist a sequence of $(2^{nR},n)$ codes with arbitrarily small probability for decoding error, i.e., $\lim\nolimits_{n\rightarrow\infty}P_r(\hat{W}\neq W)=0$, and the sensing distortion constraint $\limsup_{n\rightarrow\infty}\mathbb{E}[d(S_T^n,\hat{S}_T^n)]\le D$ is satisfied. The capacity-distortion tradeoff $C(D)$ is defined as the supremum of rate $R$ such that pair $(R,D)$ is achievable for any given $D$.
\emph{Our goal is to characterize the capacity-distortion function for the general model considered and examine its communication and sensing performance tradeoff via examples.}

\section{Optimal Capacity-Distortion Tradeoff}\label{sec:capacity_distortion_tradeoff}
In this section, we characterize the optimal capacity-distortion tradeoff $C(D)$ for the generalized 
ISAC channel model. The main result is first provided and some
properties of function $C(D)$ and the state estimator are then presented. The proof is then elaborated with focus on the converse part.
\begin{theorem}\label{the:capacity_distortion_tradeoff}
	The optimal capacity-distortion tradeoff of the point-to-point ISAC channel considered is given by 
	\begin{equation}\label{equ:theorem_result}
		C(D) = \max_{\mathcal{Q}_X} I(X;Y|S_R),\notag
	\end{equation}
	where $\mathcal{Q}_X=\{P_X:\mathbb{E}[d(S_T,\hat{S}_T)]\le D\}$ is the set of input distributions satisfying the sensing distortion constraint, and the joint distribution of variables $SS_TS_RXYZ\hat{S}_T$ is given by
	\begin{equation}\label{equ:jointPro}
	 P_X(x)P_{SS_TS_R}(ss_Ts_R)P_{YZ|XS}(yz|xs)P_{\hat{S}_T|XZ}(\hat{s}_T|xz).
	\end{equation}
\end{theorem}

\begin{remark}
	Based on the result in Theorem \ref{the:capacity_distortion_tradeoff}, there are
	\begin{enumerate}
		\item Theorem \ref{the:capacity_distortion_tradeoff} contains the result in  \cite{kobayashi2018joint} as a special case when $S_T=S_R=S$, i.e., the sensing state is the same as the channel state, and the receiver has perfect CSI.
		\item The capacity-distortion function $C(D)$ in Theorem \ref{the:capacity_distortion_tradeoff} specialized to 
		$C(D=\infty) = \max_{P_X}I(X;Y|S_R)$
		when no sensing task is required at the transmitter. 
	\end{enumerate}
\end{remark}

Before proving Theorem \ref{the:capacity_distortion_tradeoff}, the properties of $C(D)$ and the optimal state estimator are provided as follows.
\begin{lemma}\label{lem:property_C_D}
	The capacity-distortion function $C(D)$ is nondecreasing concave for $D\ge D_{min}\triangleq\min\mathbb{E}[d(S_T,\hat{S}_T)]$ where the minimum is over all $P_X$ and $P_{\hat{S}_T|XZ}$.
\end{lemma} 
\begin{lemma}\label{lem:state_estimator}
	For any given input $x$ and feedback $z$, the optimal state estimator can be chosen as a deterministic function 
	\begin{equation}\label{equ:state_estimator}
		\hat{s}_T(x,z) = \arg\min_{s'_T\in\mathcal{S}_T}\sum_{s_T\in\mathcal{S}_T}P_{S_T|XZ}(s_T|x,z)d(s_T,s'_T).\notag
	\end{equation}
\end{lemma} 

The detailed proof for Theorem \ref{the:capacity_distortion_tradeoff} are elaborated as follows. 

\subsection{Converse}
From Fano's inequality, we have 
\begin{align}
	nR& \le I(W;Y^n|S_R^n) + n\epsilon_n\notag\\
	& = \sum_{i=1}^nI(W;Y_i|Y^{i-1},S_R^n) + n\epsilon_n\notag\\
	& = \sum_{i=1}^n H(Y_i|Y^{i-1},S_R^n) - H(Y_i|Y^{i-1},S_R^n,W) + n\epsilon_n\notag\\
	& \overset{(a)}\le \sum_{i=1}^n H(Y_i|S_{R,i}) - H(Y_i|Y^{i-1},S_R^n,W) + n\epsilon_n\notag\\
	& \overset{(b)}\le \sum_{i=1}^n H(Y_i|S_{R,i}) - H(Y_i|Y^{i-1},S_R^n,W,X_i) + n\epsilon_n\notag\\
	& \overset{(c)}= \sum_{i=1}^n H(Y_i|S_{R,i}) - H(Y_i|S_{R,i},X_i) + n\epsilon_n\notag\\
	& = \sum_{i=1}^n I(X_i;Y_i|S_{R,i}) + n\epsilon_n,\notag
\end{align}
where both $(a)$ and $(b)$ follow because conditioning reduces entropy; $(c)$ follows because $(W,Y^{i-1},\{S_{R,l}\}_{l\neq i})-(S_{R,i},X_i)-Y_i$ forms a Markov chain.
Consequently, we have that
\begin{equation}
	\begin{aligned}
		R &\le \frac{1}{n}\sum_{i=1}^n I(X_i;Y_i|S_{R,i}) + \epsilon_n \\ 
		&\overset{(a)}\le \frac{1}{n}\sum_{i=1}^n C(\mathbb{E}[d(S_{T,i},\hat{S}_{T,i})]) + \epsilon_n \\ 
		&\overset{(b)}\le C\bigg(\frac{1}{n}\sum_{i=1}^n\mathbb{E}[d(S_{T,i},\hat{S}_{T,i})]\bigg) + \epsilon_n \\
		& \overset{(c)}\le C(D) + \epsilon_n, \notag
	\end{aligned}
\end{equation}
where $(a)$ follows from the definition of $C(D)$; both $(b)$ and $(c)$ follow from the concavity and nondecreasing property of $C(D)$ shown in Lemma \ref{lem:property_C_D}, respectively.

\subsection{Achievability}
The achievability proof 
uses random codebook 
and joint typically decoding, similar to that of \cite{kobayashi2018joint}. The key difference is the decoder takes into account the available CSI for decoding in addition to the received signal. The detailed proof is provided in Appendix \ref{appdedix:ache}.


\section{Numerical Algorithm for Evaluating the Optimal Capacity-Distortion Tradeoff}\label{sec:algorithm}
The optimal capacity-distortion tradeoff for the model considered is represented in form of 
an optimization problem subject to a sensing distortion constraint $\mathbb{E}[d(S_T,\hat{S}_T)]\le D$. Solving this optimization problem and determining the boundary of the capacity-distortion tradeoff region is non-trivial in general except for very few special cases. In the following, we propose a general numerical algorithm to find solutions for the tradeoff optimization, the utility of which will be further demonstrated via some examples in the next section.

\subsection{Problem Formulation}
The optimal tradeoff is an optimization problem that maximizes the mutual information over a constrained input distribution $P_X$ as in (\ref{equ:theorem_result}). For a practical system, the channel input is usually further subject to additional cost constraint such as power constraint. Accounting for this, we introduce an additional generic cost function  $b(X^n)=\frac{1}{n}\sum_{i=1}^nb(X_i)$ such that $\limsup_{n\rightarrow\infty}\mathbb{E}[b(X^n)]\le B$ into the optimization.

Then, the capacity-distortion tradeoff optimization is 
\begin{subequations}\label{opt:opti_pro_rewritten}
	\begin{align}
		\text{maximize}\quad  &\mathcal{I}(P_X;P_{Y|XS_R}|P_{S_R})\\
		\text{subject to}\quad &\sum_{x}P_X(x)b(x)\le B,\label{cons:opti_pro_penalty_form_input_cost}\\
		&\sum_{x}P_X(x)c(x)\le D,\\
		&\sum_{x}P_X(x)=1,\label{cons:opti_pro_penalty_form_pro}
	\end{align}
\end{subequations}
where the average distortion is represented by introducing an auxiliary term $c(x)$ defined as 

	\begin{align}
		c(x)=&\sum_{z}P_{Z|X}(z|x)\sum_{s_T}P_{S_T|XZ}(s_T|xz)d(s_T,\hat{s}_T(x,z)),\notag
	\end{align}
and the mutual information function for $S_R\neq\phi$ is 

\begin{table}[t]
	\begin{center} \caption{Iteration Algorithm for Problem (\ref{opt:opti_pro_penalty_form})} \label{table:algorithm}
		\resizebox{0.48\textwidth}{!}{\begin{tabular}{l}
				\hline
				{\bf Initialization}: $\mu$, the penalty parameter; $\sigma_1,\sigma_2$, convergence parameter;\\
				\quad \quad \quad \quad \quad \quad  $P_X^{(0)}(x) = \frac{1}{|\mathcal{X}|},\forall x\in\mathcal{X}$, the initial input distribution;\\
				\quad \quad \quad \quad \quad \quad  $\lambda^{(0)}$, the initial dual variable;  $k=1,l=1$, index of iteration;\\
				While $1$ do\\
				\quad 1) Update $Q_{X|YS_R}^{(k)}(x|ys_R)$ based on $P_X^{(k-1)}(x)$ and (\ref{equ:Q_update}). Set $l=1$.\\
				\quad 2) While $1$, do\\
				\quad\quad 2.1) Update $P_X^{(l)}(x)$ based on $Q_{X|YS_R}^{(k)}(x|ys_R)$, $\lambda^{(l-1)}$, (\ref{equ:PX_update}), and (\ref{equ:g_update}).\\
				\quad\quad 2.2) Update dual variables $\lambda^{(l)}$:  \\
				\quad\quad\quad\quad\quad\quad\quad$\lambda^{(l)} = \bigg[\lambda^{(l-1)} + \alpha_l\bigg(\sum_xP_X^{(l)}(x)b(x) - B\bigg)\bigg]^+ $\\
				\quad\quad 2.3) If $|\lambda^{(l)}- \lambda^{(l-1)}|\le\sigma_2$, update $P_X^{(k)}(x)=P_X^{(l)}(x),\forall x\in\mathcal{X}$, break;\\
				\quad\quad\quad\quad otherwise, $l=l+1$.\\
				\quad 3) If $||P_X^{(k)}- P_X^{(k-1)}||_2^2\le\sigma_1$, break; otherwise, $k=k+1$.\\
				{\bf Output}: $P_X^{(k)}(x),x\in\mathcal{X}$.
				\\\hline
		\end{tabular}}
	\end{center}
	\vspace{0em}
\end{table}

\begin{equation}
	\begin{aligned}
		\mathcal{I}(P_X;&P_{Y|XS_R}|P_{S_R})=\sum_{s_R}P_{S_R}(s_R)\sum_{x}\sum_{y} \\
		&P_X(x)P_{Y|XS_R}(y|xs_R)\log\frac{P_{X|YS_R}(x|ys_R)}{P_{X}(x)}. \notag
	\end{aligned}
\end{equation}
\begin{remark}
	When $S_R=\phi$, there is 
	\begin{equation}
		\begin{aligned}
			\mathcal{I}(P_X;&P_{Y|X})= \sum_{x}\sum_{y}P_X(x)P_{Y|X}(y|x)\log\frac{P_{X|Y}(x|y)}{P_{X}(x)},\notag
		\end{aligned}
	\end{equation}
	where $P_{X|Y}$ denotes the conditional PMF of $X$ given $Y$.
\end{remark}

Problem (\ref{opt:opti_pro_rewritten}) is a convex optimization problem as the mutual information function is convex and the constraints are all linear. Although the existing interior-point-method based solver such as MATLAB-CVX can be applied to find solution, it suffers high complexity for the cases with large alphabet of input $X$ and state $S$. To this end, an alternating method motivated by Blahut-Arimoto algorithm \cite{blahut1972computation} is proposed.

\subsection{Alternating solving method}\label{sec:alternating_solving_method}
Due to the page limits, in this subsection, we are focused on the general non-degenerated case $S_R\neq\phi$. The results can be easily extended to the case $S_R=\phi$.
Problem (\ref{opt:opti_pro_rewritten}) contains two cost constraints. One is related to the input cost constraint, and the other is related to the distortion constraint. To solve (\ref{opt:opti_pro_rewritten}) efficiently, we first introduce auxiliary variables $Q_{X|YS_R}(x|ys_R)=P_{X|YS_R}(x|ys_R)$ and consider incorporating the distortion constraint as a penalty term in the objective function 
\begin{equation}\label{opt:opti_pro_penalty_form}
		\max_{P_X}\max_{Q_{X|YS_R}} \mathcal{J}(P_X,Q_{X|YS_R}) - \mu\sum_{x}P_X(x)c(x)
\end{equation}
subject to constraints (\ref{cons:opti_pro_penalty_form_input_cost}) and (\ref{cons:opti_pro_penalty_form_pro}),
where 
\begin{align}\label{opt:obj_fun_mut_fun}
	\mathcal{J}(P_X,&Q_{X|YS_R}) =\sum_{s_R}P_{S_R}(s_R)\sum_{x}\sum_{y}\notag\\
	&P_X(x)P_{Y|XS_R}(y|xs_R)\log\frac{Q_{X|YS_R}(x|ys_R)}{P_{X}(x)}, \notag
\end{align}
and $\mu\ge0$ is a fixed parameter. By varing $\mu$, we obtain the capacity-distortion tradeoff for fixed cost constraint of input. 
\begin{proposition}
	(\ref{opt:opti_pro_penalty_form}) can be solved iteratively in closed-form. 
	\begin{enumerate}
		\item For fixed $P_X$, the optimal value of $Q_{X|YS_R}$ for (\ref{opt:opti_pro_penalty_form}) is given in-closed form as 
		
		\begin{equation}\label{equ:Q_update}
			\begin{aligned}
				Q&_{X|YS_R}^*(x|ys_R) 
				=\frac{P_X(x)P_{Y|XS_R}(y|xs_R)}{\sum\nolimits_{x'}P_X(x')P_{Y|XS_R}(y|x's_R)}.
			\end{aligned}
		\end{equation}
		\item For fixed $Q_{X|YS_R}$, the optimal value of $P_X$ is given as 
		\begin{equation}\label{equ:PX_update}
			\begin{aligned}
				P_X(x)^* = \frac{2^{g(x)}}{\sum_{x'}2^{g(x)}},
			\end{aligned}
		\end{equation}
		where 
		\begin{equation}\label{equ:g_update}
			\begin{aligned}
				g(x) =& \sum_{s_R}\sum_{y}P_{S_R}(s_R)P_{Y|XS_R}(y|xs_R)\\
				&\log Q_{X|YS_R}(x|ys_R) - \lambda^* b(x) - \mu c(x),
			\end{aligned}
		\end{equation}
		and $\lambda^*$ is the optimal dual variable for constraint (\ref{cons:opti_pro_penalty_form_input_cost}).
	\end{enumerate}
\end{proposition}

The detailed algorithm is shown in Table \ref{table:algorithm}. As mentioned, such an algorithm yields an input distribution $P_{X}$ corresponding to a pair of capacity and distortion values $(C_{\mu},D_{\mu})$. By varing $\mu$, we obtain the capacity-distortion tradeoff for the given channel model under the input cost constraint.

\section{Examples}\label{sec:example}
In this section, two non-trivial examples are constructed to illustrate the benefits of ISAC with respect to the separation-based approach over generalized memoryless point-to-point channels. 
Here the separation-based approach refers to a scheme where orthogonal resources are divided into either pure state sensing via echo feedback or pure data communciation.

\subsection{Capacity-Distortion Tradeoff over Binary Channel}
Consider a binary channel
$Y = SX\mod 2$,
where both input $X$ and output $Y$ are binary distributed, while the channel state $S\in\{0,1,2,3\}$ with PMF $P_S(0)=0.1$, $P_S(1)=0.2$, $P_S(2)=0.3$, and $P_S(3)=0.4$. The sensing signal $Z$ is assumed to coincide with $Y$, i.e., $Z=Y$, similar to that of \cite{kobayashi2018joint}. The sensing state $S_T$ is assumed as a function of channel state where $S_T=0$ if $S=0,1,2$ while $S_T=1$ if $S=3$. The Hamming distortion measure $d(s_T,\hat{s}_T)=s_T\oplus\hat{s}_T$ is considered. Two different instances of CSIR are considered:
\begin{enumerate}
	\item $S_R=\phi$, the receiver has no CSI.
	\item $S_R=S$, the receiver has perfect CSI.
\end{enumerate}
Next, we characterize 
$P_X(0)\triangleq p$ that maximize $C(D)$. 
\begin{proposition}
	The capacity-distortion tradeoff of the binary channel $Y = SX\mod 2$ is given by
	\begin{enumerate}
		\item when $S_R=\phi$, $C(p)=H_2(0.6-0.6p)-(1-p)H_2(0.4)$, $D(p)=0.2(1+p)$;
		\item when $S_R=S$, $C(p)=0.6H_2(p)$,  $D(p)=0.2(1+p)$.
	\end{enumerate}
	where $H_2(p)$ denotes the binary entropy function.
\end{proposition}
\begin{IEEEproof}
	\begin{enumerate}
		\item when $S_R=\phi$, the capacity is given as 
		\begin{equation}
			\begin{aligned}
				C(p) &= H(Y) - H(Y|X),\notag
			\end{aligned}
		\end{equation}
		where the output $Y$ is Bernoulli distributed such that $P_Y(1) = 0.6(1-p)$, and
		\begin{align}\label{equ:condi_entro_Y_X}
				H(Y|X) &
				= (1-p)H_2(0.4),
		\end{align}
		where the  equality in (\ref{equ:condi_entro_Y_X}) follows because $P_{Y|X}(0|0)=1$ and $P_{Y|X}(0|1)=0.4$.
		\item when $S_R=S$, the capacity is given by
		\begin{equation}
			\begin{aligned}
				C(p) = H(Y|S)=0.6H_2(p),\notag
			\end{aligned}
		\end{equation}
		where the last equality follows because $P_{Y|S}(0|0)=1$, $P_{Y|S}(0|2)=1$, $P_{Y|S}(0|1)=p$, and $P_{Y|S}(0|3)=p$.
	\end{enumerate}
	The expected distortion functions for the considered two cases are the same. We first determine the state estimator $\hat{s}_T(x,z)$ by Lemma \ref{lem:state_estimator} as follows:
	\begin{equation}\label{equ:state_estimator_binary}
		\begin{aligned}
			&\hat{s}_T(x,0) = 0, \forall x, \quad\hat{s}_T(x,1) = 1, \forall x.
		\end{aligned}
	\end{equation}
	Then, the expected distortion over all input $X$ and feedback $Z$ based on the state estimator in (\ref{equ:state_estimator_binary}) is given by 
	\begin{align}
			\mathbb{E}[d(S_T,&\hat{S}_T)]
			 = 0.2(1+p),\notag
	\end{align}
	which completes the proof.
\end{IEEEproof}

In particular, we consider two extreme points for each case.
\begin{enumerate}
	\item $S_R=\phi$: If $p=0$, i.e., the encoder always sends $X=1$, the minimum distortion $D_{min}=0.2$ is achieved, and the corresponding capacity $C(D_{min})=0$. If $p=0.5905$, the maximum capacity $C_{max}=0.4068$ is achieved, and the corresponding distortion is $D=0.3181$.
	\item $S_R=S$: If $p=0$, i.e., the encoder always sends $X=1$, the minimum distortion $D_{min}=0.2$ is achieved, and the corresponding capacity $C(D_{min})=0$. If $p=0.5$, the maximum capacity $C_{max}=0.6$ is achieved, and the corresponding distortion is $D=0.3$.
\end{enumerate}

The results are shown in Fig \ref{fig:CDA_SR_phi}, where TSA denotes the separation-based approach  achieved via a time-sharing between capacity-distortion pair $(C,D)=(0,0.2)$ and $(0.4068,0.4)$ for $S_R=\phi$ and between $(0,0.2)$ and $(0.6,0.4)$ for $S_R=S$, respectively. 
It is noted that this upper extreme distortion $D=0.4$ can be trivially achieved by considering a fixed estimator $\hat{s}_T = 0$ which thus imposes no extra constraint on the communication input and incurs zero loss of capacity. 

\subsection{Capacity-Distortion Tradeoff over Real Gaussian Channel}\label{sec:GaussianChannelPerfectFeedback}

\begin{figure}[!t]
	\centering
	\includegraphics[width=0.9\linewidth]{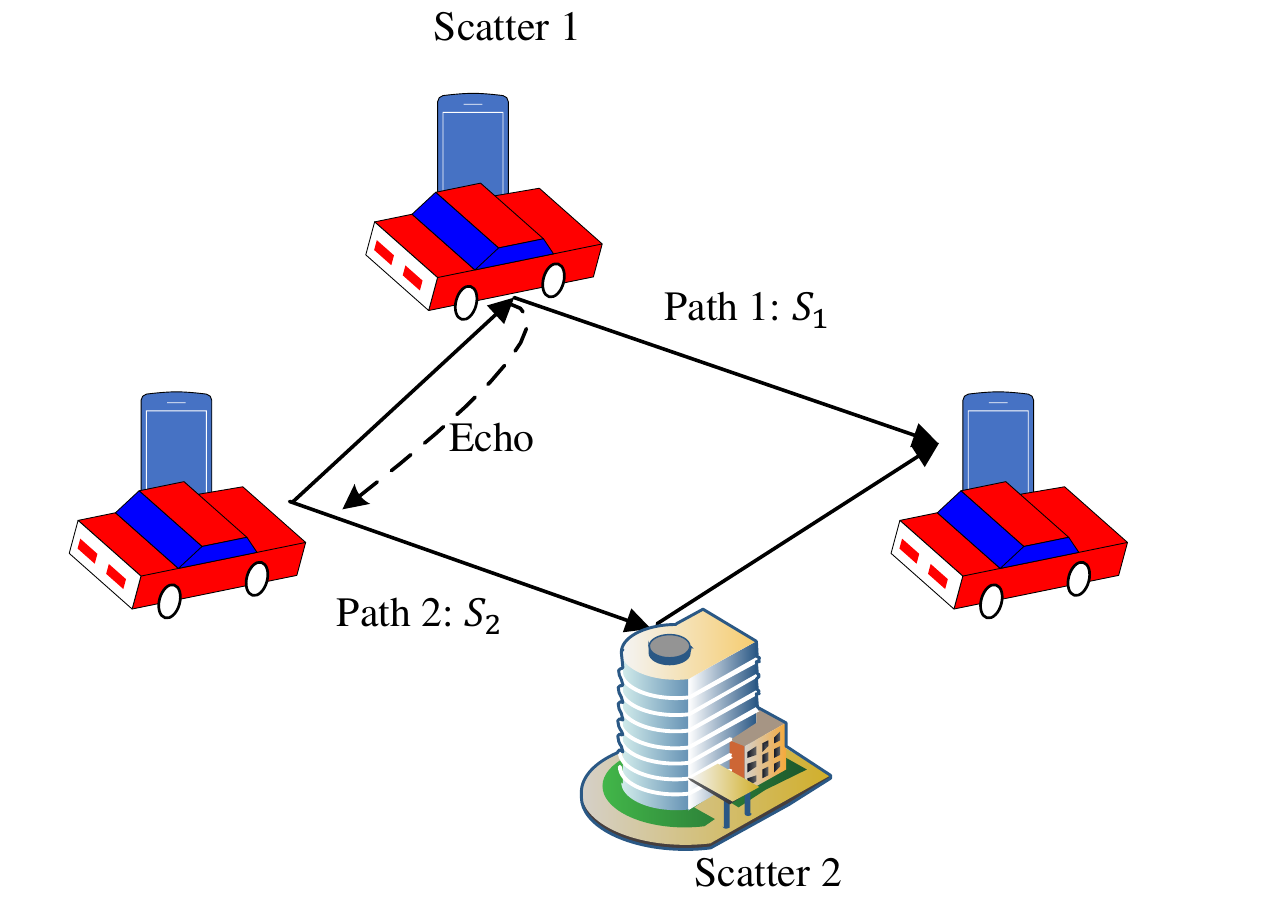}
	\caption{Illustration of ISAC over real Gaussian channel in vehicular networks.}
	\label{fig:exampleIllu}
	\vspace{0em}
\end{figure}

In the second example, we consider that the sensing target is part of the scatters in the surrounding environment of the car, which is shown in Fig. \ref{fig:exampleIllu}. The communication channel is a real Gaussian channel model with Rayleigh fading, i.e.,
\begin{align}
	Y_i = S_iX_i + N_i, \notag
\end{align}
where the input $X_i$ is subject to the power constraint $\frac{1}{n}\sum_i\mathbb{E}[|X_i|^2]\le P$, and the channel contains two path, i.e., $S_i=S_{1,i}+S_{2,i}$. State components $S_{1,i}$, $S_{2,i}$ and noise $N_i$ are i.i.d. Gaussian distributed with zero mean and unit variance.  A noisy sensing echo signal 
\begin{align}
	Z_i = \alpha S_{1,i}X_i+ N_{\text{fb},i} \notag
\end{align}
is considered,
where $\alpha$ is the reflection coefficient, 
and $\{N_{\text{fb},i}\}$ are i.i.d. Gaussian distributed with zero mean and unit variance.
The sensing state $S_T$ is $S_{T,i}=\alpha S_{1,i}+V_i$, where $V_i$ denotes the noise of sensing that is i.i.d. Gaussian distributed with zero mean and unit variance. The quadratic distortion measure $d(s_T,\hat{s}_T)=(s_T-\hat{s}_T)^2$ is considered. Two cases of CSIR $S_R=\phi$ and $S_R=S$ are considered. 

For this continuous channel, calculating the capacity-distortion tradeoff means that finding the optimal distribution of input $X$, which is hard to be obtained. Thus, discretization method is applied to obtain a numerical approximation for the capacity-distortion tradeoff. We consider a fixed input power constraint $P=10$ dB, a fixed reflection coefficient $\alpha=0.5$, and the input $X$ is quantized to a Pulse-Amplitude-Modulation (PAM) like constellation $\mathcal{X}_q=[-10:q:10]$ with step size $q$ to be determined. The channel states $S_1$ and $S_2$ are quantized with alphabet $[-5:q:5]$. The communication noise $N$ is quantized with alphabet $[-5:q^2:5]$. The noise in echo signal $N_{\text{fb}}$ is quantized with alphabet $[-5:\alpha q^2:5]$. The sensing noise $V$ is quantized with alphabet $[-5:\alpha q:5]$.
Denoting the resultant quantized input, noise, and channel state by $X_q,N_q,N_{\text{fb},q},V_q,S_q$, the channel output and sensing state become $Y_q=S_qX_q+N_q$, $Z_q=\alpha S_{1,q}X_q+N_{\text{fb},q}$, and $S_{T,q}=\alpha S_{1,q}+V_q$. Then,
the alternating solving method proposed in Sec. \ref{sec:alternating_solving_method} is applied to obtain the optimal input.

\begin{figure*}
	\begin{minipage}[t]{0.325\linewidth}
		\vspace{0pt}
		\centering
		\includegraphics[width=1\linewidth]{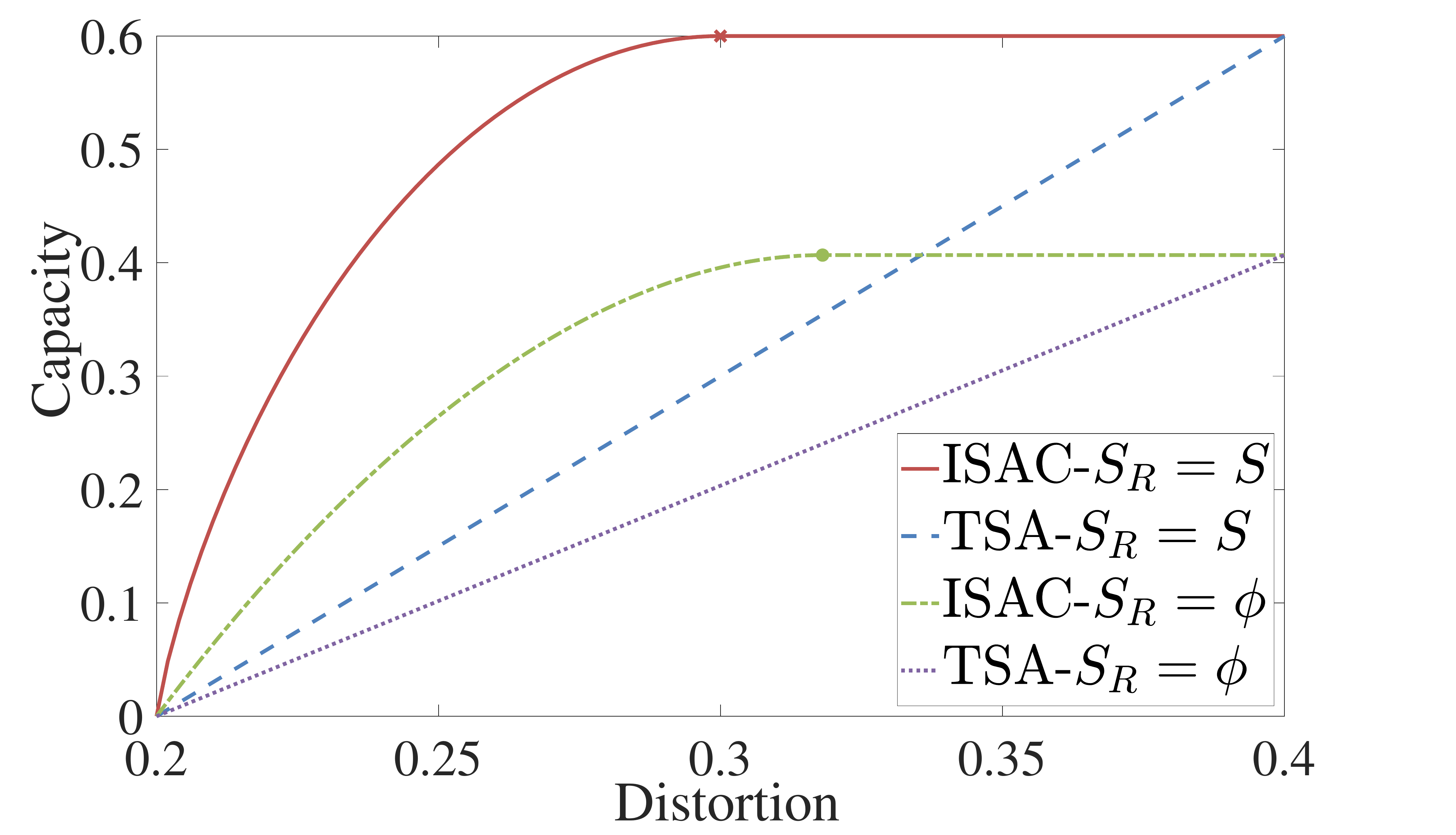}\\
		\caption{Capacity-distortion tradeoff of binary channel.}
		\label{fig:CDA_SR_phi}
	\end{minipage}
	\hspace{.01in}
	\begin{minipage}[t]{0.325\linewidth}
		\vspace{0pt}
		\centering
		\includegraphics[width=1\linewidth]{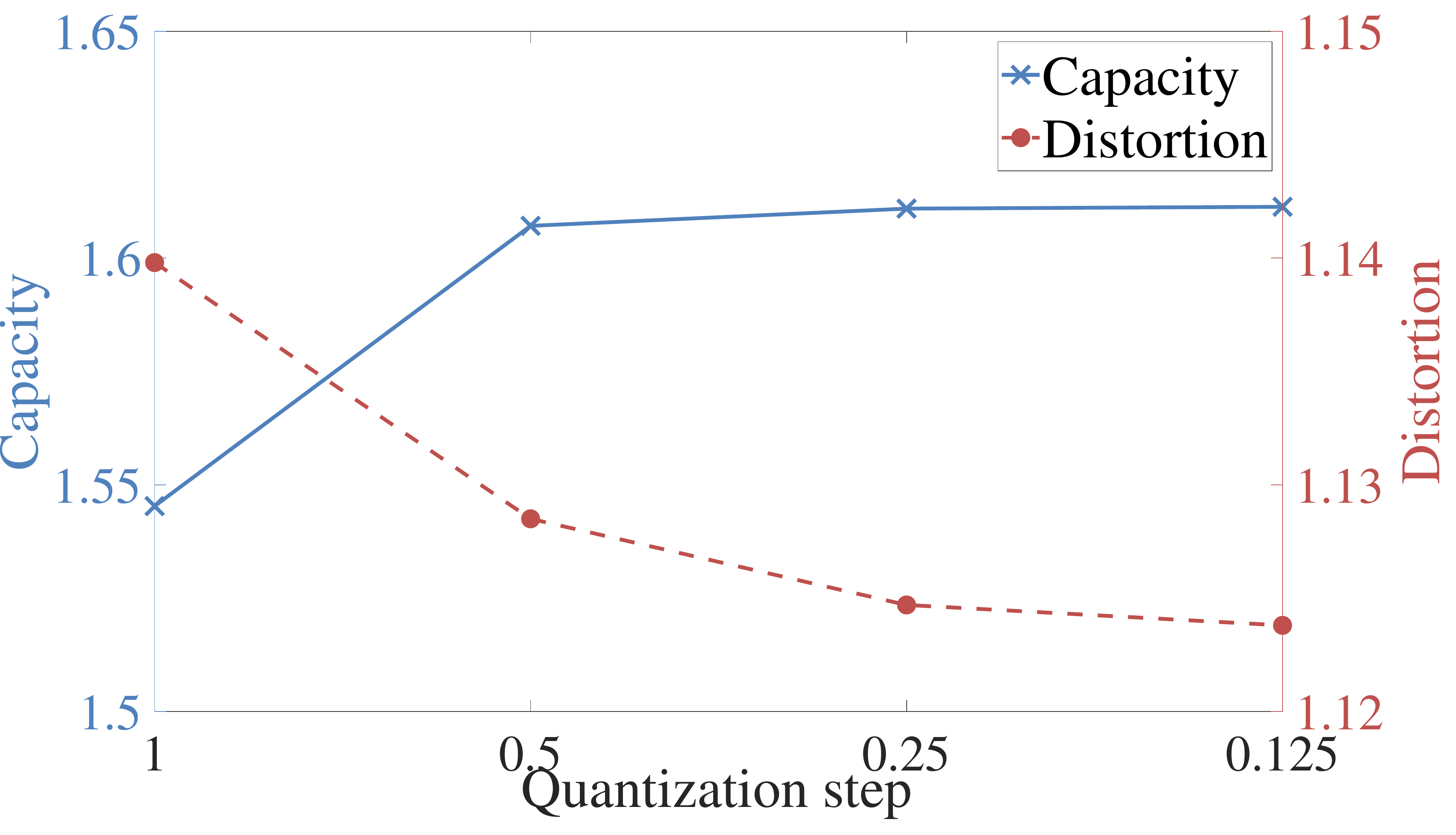}\\
		\caption{The impact of quantization step on the value of capacity-distortion pair.}
		\label{fig:impactQuan}
	\end{minipage}
	\hspace{.01in}
	\begin{minipage}[t]{0.325\linewidth}
		\vspace{0pt}
		\centering
		\includegraphics[width=1\linewidth]{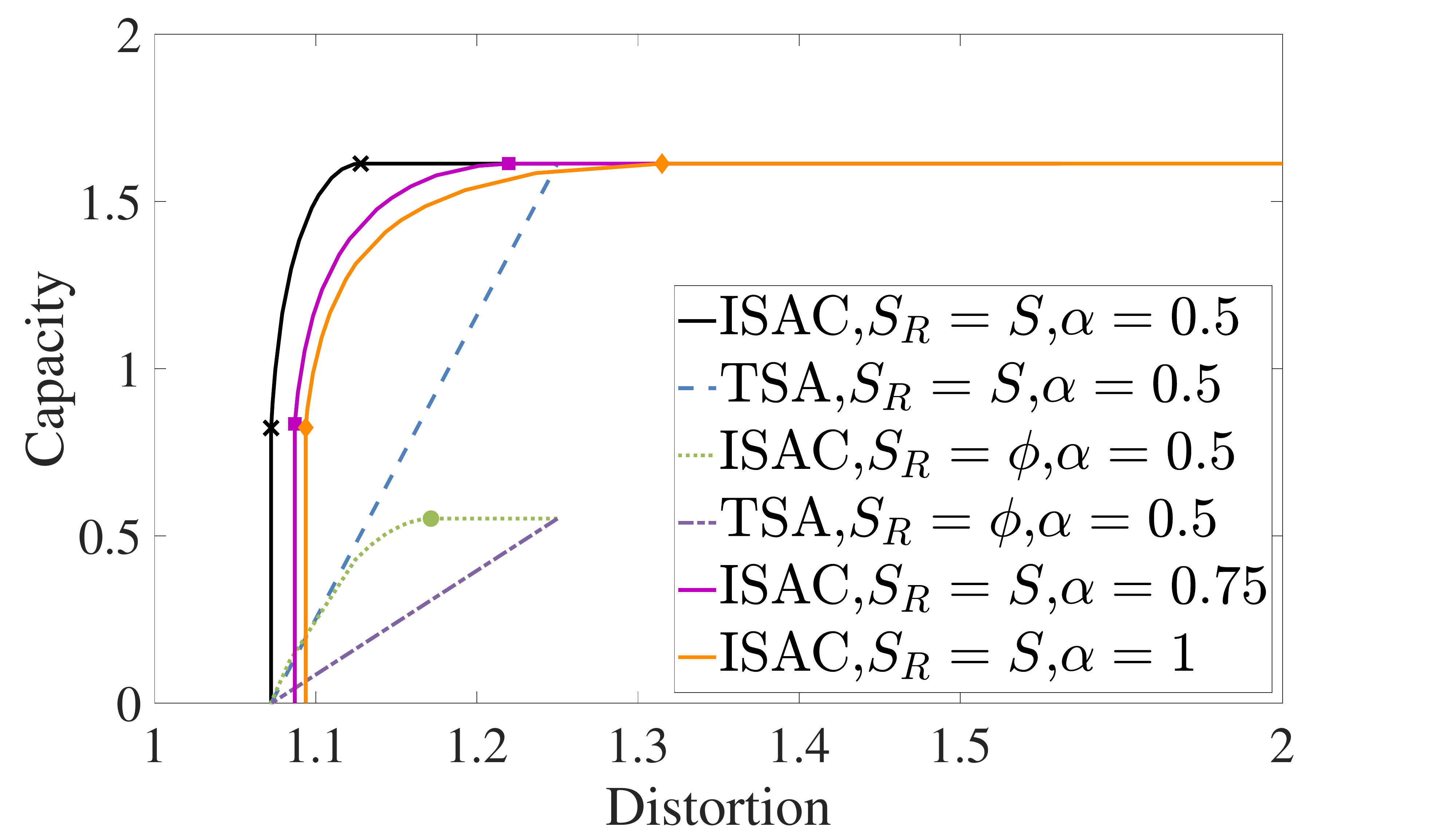}\\
		\caption{Capacity-distortion tradeoff for real Gaussian channel with noisy sensing echo signal.}
		\label{fig:GaussianChaImperfectFeedback}
	\end{minipage}
\end{figure*}

To determine how small $q$ should be sufficient for the computation, we first evaluate the impact of quantization step $q$ on the value of capacity-distortion pair $(C,D)$. More specifically, we consider the case $S_R=S$, and the penalty parameter $\mu$ is set as $1$ in the numerical algorithm. The set of values for quantization step $q$ is set as $\{1,0.5,0.25,0.125\}$. The results are shown in Fig. \ref{fig:impactQuan}. It can be found that $q=0.125$ suffices to achieve a quite satisfactory approximation. Thus, in the following numerical evaluations, $q$ is set as $0.125$.

We compare ISAC with TSA when $\alpha=0.5$ for both $S_R=\phi$ and $S_R=S$. We also test the impact of reflection coefficient with the consideration of $\alpha=0.75$ and $\alpha=1$ for $S_R=S$. When $\alpha=0.5$, there are two extreme points for each case.
\begin{enumerate}
	\item $S_R=\phi$: On one extreme, the minimum distortion $D_{min}$ is achieved by 2-ary PAM and is equal to $1+\frac{\alpha^2}{1+\alpha^2P}=1.0714$, while corresponding capacity $C(D_{min})=0$. On the other extreme, the maximum capacity $C_{max}=0.5514$ is achieved by applying the alternating solving method, and the corresponding distortion is $D=1.1719$.
	\item $S_R=S$: On one extreme, the minimum distortion $D_{min}$ is still $1.0714$, 
	and the corresponding capacity becomes $C(D_{min})=0.8223$. On the other extreme, the maximum capacity $C_{max}=\frac{1}{2}\mathbb{E}_{S}[\log(1+S^2P)]=1.6128$ is achieved when $X$ is Gaussian distributed with zero mean and variance $P$, while the corresponding distortion is $D=1 + \mathbb{E}_{X}[\frac{\alpha^2}{1+\alpha^2X^2}]=1.1279$.
\end{enumerate}

The entire capacity-distortion regions are shown in Fig \ref{fig:GaussianChaImperfectFeedback}. The results of TSA are achieved via a time-sharing between pair $(C,D)=(0,1.0714)$ and $(0.5514,1.25)$ for $S_R=\phi$ and between $(0,1.0714)$ and $(1.6128,1.25)$ for $S_R=S$ when $\alpha=0.5$, respectively. The upper extreme distortion $D=\text{var}[S_T]=1.25$ is achieved by considering a fixed estimator $\hat{s}_T = 0$ . The results reveal that ISAC can indeed provide a significant gain over the separation-based approach. 
Besides, with the increase of $\alpha$, the capacity-distortion region becomes smaller. The reason is that the value of $\alpha$ affects the SNR of echo signal, the sensing state, and then the performance of sensing, while the communication rate keeps constant.

\section{Conclusion}\label{sec:conclusion}
In this work, we have investigated the fundamental limit of ISAC over general memoryless point-to-point channels. A new state-dependent ISAC channel model with correlated sensing state and channel state, and imperfect CSIR is formulated. Based on the model, the optimal capacity-distortion tradeoff is characterized. Two examples are further constructed to demonstrate that ISAC provides a significant gain for both communication and sensing as compared to the separation-based approach. As future work, it is of great interest to extend the current framework to multi-terminal ISAC topologies (such as MACs and BCs), establishing the fundamental limits to support the efficient application of ISAC in vehicular networks.

\appendices
\section{Achievability Proof of Theorem \ref{the:capacity_distortion_tradeoff}}\label{appdedix:ache}
In the achievability proof, we show that the capacity-distortion tradeoff in Theorem \ref{the:capacity_distortion_tradeoff} is achieved when the distortion function $d(\cdot)$ is bounded by $d_{max} = max_{(s_T,\hat{s}_T)\in\mathcal{S}_T\times\hat{\mathcal{S}}_T}d(s_T,\hat{s}_T)<\infty$. The achievable coding scheme is given as follows. 
\begin{itemize}
	\item \emph{Codedbook generation}: Fix the input distribution $P_X(\cdot)$ and state estimating function $\hat{s}_T(x,z)$ that achieves $C(D/(1+\epsilon))$, where $D$ is the desired distortion. Randomly and independently generate $2^{nR}$ sequences $x^n(w)$ for each message $w\in[1:2^{nR}]$. The collection of these sequences form a codedbook $\mathcal{C}$ which is then revealed to both the transmitter and receiver.
	\item \emph{Encoding}: To send a message $w\in[1:2^{nR}]$, the encoder chooses and transmits the $w$th codeword $x^n(w)$.
	\item \emph{Decoding}: The decoder finds a unique message $\hat{w}$ such that $(y^n,s_R^n,x^n(\hat{w}))$ is jointly typical, i.e.,
	\begin{equation}
		(y^n,s_R^n,x^n(\hat{w})) \in \mathcal{T}_\epsilon^{(n)}. \notag
	\end{equation}
	\item \emph{Estimation}: The encoder computes the resconstruction sequence for sensing state based on the state estimating functions as $\hat{s}_T^n = \hat{s}_T(x^n(w),z^n)$.
\end{itemize}

The analysis of the probability of decoding error and the expected distortion is given as follows. 

\emph{Analysis of the probability of decoding error}: Due to the symmetry of the random codebook generation in the proposed scheme, the probability of decoding error averaged over all messages and codebooks $P_e$ is equal to the the probability of decoding error when a certain message is transmitted. Thus, we assume without loss of generality that message $w=1$ is sent, the decoder makes an error if and only if one or both of the following events occur: 
\begin{equation}
	\mathcal{E}_1 = \{(y^n,s_R^n,x^n(1))\notin\mathcal{T}_\epsilon^{(n)}\}, \notag
\end{equation}
\begin{equation}
	\mathcal{E}_2 = \{(y^n,s_R^n,x^n(\hat{w})) \in \mathcal{T}_\epsilon^{(n)}\enspace\text{for some} \enspace w\neq1\}.\notag
\end{equation}
By the union of events bound, 
\begin{equation}
	P_e = P(\mathcal{E}_1 \cup\mathcal{E}_2) \le P(\mathcal{E}_1) + P(\mathcal{E}_2).\notag
\end{equation}
By the law of large numbers, the first term $P(\mathcal{E}_1)$ tends to zero as $n\rightarrow\infty$. By the independence of the codebooks and the packing lemma \cite{el2011network}, the second term tends to zero as $n\rightarrow\infty$ if $R<I(X;Y|S_R)$. Therefore, the probability of decoding error $P_e$ tends to zeros as $n\rightarrow\infty$ if $R<I(X;Y|S_R)$.

\emph{Analysis of expected distortion}: The expected distortion averaged over the random codebook, encoding, and decoding, is upper bounded as
\begin{equation}
	\begin{aligned}
		&\limsup_{n\rightarrow\infty}\mathbb{E}[d(S_T^n,\hat{S}_T^n)]\\
		&\overset{(a)}\le \limsup_{n\rightarrow\infty}\bigg(P_ed_{max}+(1-P_e)(1+\epsilon)\mathbb{E}[d(S_T,\hat{S}_T)]\bigg)\\
		&\overset{(b)}\le \limsup_{n\rightarrow\infty}(P_ed_{max}+(1-P_e)(1+\epsilon)\frac{D}{1+\epsilon})\\
		&\overset{(c)}=D,\notag
	\end{aligned}
\end{equation}
where $(a)$ follows by applying the upper bound of the distortion function to the decoding error event and the typical average lemma \cite{el2011network} to the successful decoding event; $(b)$ follows from the random codebook generation with the input distribution $P_X(\cdot)$ and state estimating function $\hat{s}_T(x,z)$ that achieves $C(D/(1+\epsilon))$; $(c)$ follows because $P_e$ tends to zeros as $n\rightarrow\infty$ if $R<I(X;Y|S_R)$. 

The above analysis shows that the capatiy-distortion tradeoff $(C(\frac{D}{1+\epsilon}),D)$ can be achieved. Therefore, by the continuity of function $C(D)$, $C(D)$ as defined in Theorem \ref{the:capacity_distortion_tradeoff} is achieved as $\epsilon\rightarrow0$.

\bibliographystyle{IEEETran}
\bibliography{ref}
\end{document}